\documentclass[%
 reprint,
 amsmath,amssymb,
 aps,
]{revtex4-2}

\usepackage{graphicx}
\usepackage{dcolumn}
 \usepackage{xspace}
\usepackage{bm}
\usepackage{lineno}
\usepackage{ragged2e}
\usepackage{float}
\usepackage{pifont}
\usepackage[utf8]{inputenc}
\usepackage[T1]{fontenc}
\usepackage{booktabs, array, mathptmx, float, tabularx, booktabs, lipsum, amsmath,multirow}
\usepackage{siunitx, xcolor}
\usepackage[version=4]{mhchem}
\graphicspath{{figs/}{figsgaoerb/}} 
\usepackage[colorlinks,linkcolor=blue,anchorcolor=blue,citecolor=blue]{hyperref}
\usepackage{graphicx}

\usepackage{color}

\begin{document}

\preprint{APS/123-QED}

\title{Coherent perfect absorber and laser induced by directional emissions in the non-Hermitian photonic crystals}

\author{{Zhifeng Li}\textsuperscript{1}}

\author{{Hai Lin}\textsuperscript{1}}

\email{Corresponding author: linhai@mail.ccnu.edu.cn}

\author{{Rongxin Tang}\textsuperscript{2}}

\email{Corresponding author: rongxint@ncu.edu.cn}

 \author{{Haitao Chen}\textsuperscript{3}}
 \author{{Jiaru Tang}\textsuperscript{1}}
 \author{{Rui Zhou}\textsuperscript{1}}
 \author{{Jing Jin}\textsuperscript{1}}
 \author{{Y. Liu}\textsuperscript{4}}

\affiliation{$^{1}$ College of Physics Science and Technology, Central China Normal University, Wuhan 430079, Hubei Province \\}

\affiliation{$^{2}$ Institute of Space Science and Technology, Nanchang University, Jiangxi 330021, Jiangxi Province \\}

\affiliation{$^{3}$ Wuhan Maritime Communication Research Institute,  Wuhan 430205, Hubei Province \\}

\affiliation{$^{4}$ School of Physics, Hubei University, Wuhan 430079, Hubei Province \\}

\begin{abstract}
In this study, we propose the application of non-Hermitian photonic crystals (PCs) with anisotropic emissions. Unlike a ring of exceptional points (EPs)  in isotropic non-Hermitian PCs, the EPs of anisotropic non-Hermitian PCs appear as lines symmetrical about the $\Gamma$ point. The non-Hermitian Hamiltonian indicates that the formation of EPs is related to the non-Hermitian strength. The real spectrum appears in the $\Gamma$Y direction and has been validated as the complex conjugate medium (CCM) by effective medium theory (EMT). But for the $\Gamma$X direction, EMT indicates that the effective refractive index has a large imaginary part, which forms an evanescent wave inside the PCs. Thence, coherent perfect absorber (CPA) and laser effects can be achieved in the directional emission of the $\Gamma$Y. The outgoing wave in the $\Gamma$X direction is weak, which can significantly reduce the losses and electromagnetic interference caused by the leakage waves. Furthermore, the non-Hermitian PCs enable many fascinating applications such as signal amplification, collimation, and angle sensors. 

\end{abstract}                              
\maketitle
\section*{Introduction} 
The Dirac cone underlies many unique electronic properties of graphene and topological insulators\cite{P_Diracquantum1928,mei_First2012,neto_electronic2009,plihal1991photonic}. However, Dirac cone dispersions are not limited to graphene but can also be found in classical wave periodic systems such as photonic crystals\cite{zhang2008observing,ochiai2010topological,li2022low,Topological2021Zhou,2022HigherZhou}. In 1987, Yablonovitch and John independently proposed the concept of photonic crystals composed of materials with different refractive indices arranged periodically\cite{yablonovitch1987inhibited,john1987strong}. In recent years, photonic crystals with Dirac dispersion have shown novel physical properties. Huang et al. have achieved a double zero-index medium (the effective permittivity and permeability are near zero) by designing a Dirac cone in C4-symmetric PCs\cite{huang2011dirac}. Since effective medium parameters approach zero for every propagation direction, PCs with Dirac dispersion behave like an isotropic near-zero index medium. The near zero-index PCs mean that the impedance can be tuned to match that of the external medium \cite{li2021dirac} and have interesting optical applications, including electromagnetic wave manipulation\cite{fleury2014manipulation}, beam focusing\cite{he2016realization}, and electromagnetic cloak\cite{hao2010super}. Besides, it was noted that Dirac cone dispersion can change with lattice symmetry. He et al. have realized semi-Dirac dispersion in rhombic lattice and demonstrated that PCs behave as an anisotropic zero medium\cite{he2015dirac}, solving the directional radiation of waves in zero-index PCs. Yasa et al. have demonstrated the formation of semi-Dirac dispersion in microwave experiments, and the transmission efficiency of incident waves in different directions vary greatly in anisotropic PCs\cite{yasa2018full}. Note that these research works are based on the Hermite system without material loss or radiation leakage. 

Over the past two decades, open systems that are described by a non-Hermitian Hamiltonian have become a subject of intense research\cite{feng2017non,el2018non,xu2021non,coppolaro2020non}.These non-Hermitian systems encompass classical wave systems with balanced gain and loss, semiclassical models with mode selective losses, and minimal quantum systems. Besides, the non-Hermitian systems exhibit a rich array of novel phenomena, such as non-Hermitian skin effects, nonreciprocity, and topological transitions. In the recent years, it was suggested that some non-Hermitian perturbations (gain or loss) could deform a Dirac cone and spawn a ring of exceptional points\cite{zhen2015spawning}. For the two-dimensional (2D) PCs, a ring of EPs can be achieved by introducing complex permittivity, whose imaginary part means loss or gain\cite{luo2021non}. In the PT-symmetric systems, the complex permittivity exhibits a symmetric spatial profile of $\varepsilon(x)=\varepsilon^*(-x)$\cite{ruter2010observation,longhi2009bloch}. The scattering of electromagnetic waves by boundaries can produce CPA and laser effects\cite{gu2021acoustic}. When the amplitude and phase of the two incident waves are well-tuned (in-phase excitation), the outgoing waves will disappear, and the resulting may lead to perfect absorption. On the contrary (out-of-phase excitation), the outgoing waves by pumping the incident waves are greatly enhanced, which can be regarded as a laser mode. Such non-Hermite PCs behave like the CCM\cite{dragoman2011complex}. A CCM was first proposed by Bai \textit{et al}.\cite{bai2016simultaneous}, who used a lattice of core-shell cylinders made of loss and gain materials, respectively. In addition, Cui \textit{et al}. showed that CCM in a wide frequency range could be realized using non-PT symmetric PCs with a ring of EPs\cite{cui2020realization}. However, PCs with a ring of EPs correspond to an isotropic CCM, supporting the waves to continue propagating in all directions. To limit wave propagation along undesirable directions, the addition of confining metallic materials should be considered. Otherwise the loss and electromagnetic interference will be caused by the leaky waves. Nevertheless, the above-discussed limitations may complicate the fabrication and utilization of non-Hermite PCs.

In this paper, we design a non-Hermitian PC with anisotropic emissions. We reveale the unique characteristics of wave propagation in this PC, including the significant amplification of the signal and collimation\cite{sun2012loss}, and demonstrate some typical applications, such as angle sensors\cite{neil2021two}, coherent perfect absorbers\cite{chong2010coherent,wu2021broadband}, and lasers\cite{hodaei2014parity}. The non-Hermitian Hamiltonian model indicates that the real spectrum appears in the $\Gamma$Y direction, while a complex spectrum appears in the $\Gamma$X direction. From the EMT viewpoint\cite{lu2007calculation}, the effective refractive index is real for the $\Gamma$Y direction, and PC behaves as a homogeneous CCM, which supports CPA and laser effects. However, for the $\Gamma$X direction, EMT indicates that the effective refractive index has a large imaginary part, which forms an evanescent wave inside the PCs\cite{shaposhnikov2022effective}. Numerical simulation results confirm that CPA and laser effects can be achieved in the directional emission of the $\Gamma$Y. The outgoing wave in the $\Gamma$X direction is weak, which can significantly reduce the losses and electromagnetic interference caused by the leakage waves. Significantly, our design without confining metallic materials will expand the application of non-Hermitian PCs in photonic devices.

\section*{The non-Hermitian Hamiltonian model} 
\begin{figure}[!ht]
\centering
\includegraphics[width=0.4\textwidth]{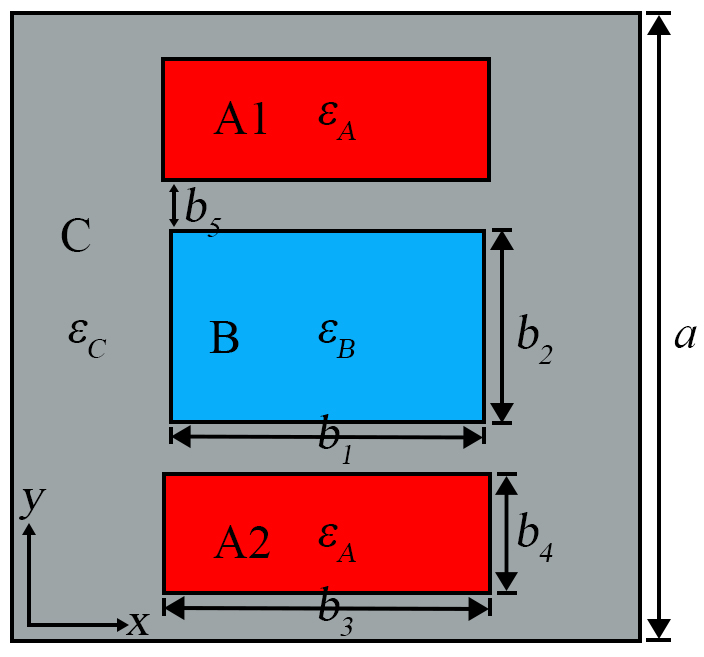}

\caption{The representation of a C2-symmetric unit cell consists of three rectangular dielectric rods with structural parameters $b_1 = 0.5a$, $b_2 = 0.3a$, $b_3 = 0.523a$, $b_4 = 0.19a$, and $b_5 = 0.085a$. The relative permittivity parameters used are $\varepsilon_A=10.04+i \gamma$, $\varepsilon_B=10.04+i \ell_r \gamma$ and $\varepsilon_C=1$. $\gamma$ represents that the rectangular area consists of a loss or (gain) medium and $\ell_r$ is defined as the loss-gain ratio.}\label{Fig1}
\end{figure}

The structure of our design is shown in \textcolor{blue}{Fig.~\ref{Fig1}}, a two-dimensional PC unit cell composed of a C2-symmetric rectangular dielectric rod. The relative permittivity of the rectangular dielectric rod (domain A1 and A2, the red region) and the rectangular dielectric rod (domain B, the blue region) are  $\varepsilon_A$ and $\varepsilon_B$, respectively. The air (domain C, the gray region) with relative permittivity $\varepsilon_C$, and the relative permeability $\mu$ of all media is 1.0. The structural parameters are $b_1 = 0.5a$, $b_2 = 0.3a$, $b_3 = 0.523a$, $b_4 = 0.19a$, and $b_5 = 0.085a$, where “$a$” is the lattice constant of the unit cell. To construct a non-Hermitian photonic crystal, we introduce the imaginary part of the relative permittivity to represent a certain amount of loss or gain. Because of the rotational symmetry of the PC (the rectangular rod has the same appearance when rotated 90° around the center point of the unit cell), the relative permittivity of the rectangular rod areas takes the form: $\varepsilon_A=10.04+i \gamma$, $\varepsilon_B=10.04+i \ell_r \gamma$. The real part of the relative permittivity is 10.04, and the positive (negative) sign of $\gamma$ indicates that the rectangular area consists of a loss or (gain) medium. The remaining air portion does not add non-Hermitian perturbation and its relative permittivity $\varepsilon_C=1$. Among them, $\ell_r$ is defined as the loss-gain ratio, and the positive sign of $\ell_r$ indicates that the red and blue regions are either both loss or gain medium. In contrast, a negative sign means that one region is loss and the other is gain. 
In principle, the eigenfrequencies and eigenfields of non-Hermitian PC can be obtained by the Helmholtz equation. However, we can construct a non-Hermitian Hamiltonian using the Bloch states of the Hermitian system as the bases\cite{wang2020effective}, and effective eigenequation can be obtained (see details in Appendix A)
\begin{equation}
\left(\boldsymbol{H}_1\right)^{-1} \cdot \boldsymbol{H}_2 \cdot \psi=\boldsymbol{H} \cdot \psi=\left(\frac{\omega_{\boldsymbol{k} n}}{c}\right)^2 \psi, \label{1}
\end{equation}
\begin{figure*}[ht]
\includegraphics[width=1\textwidth]{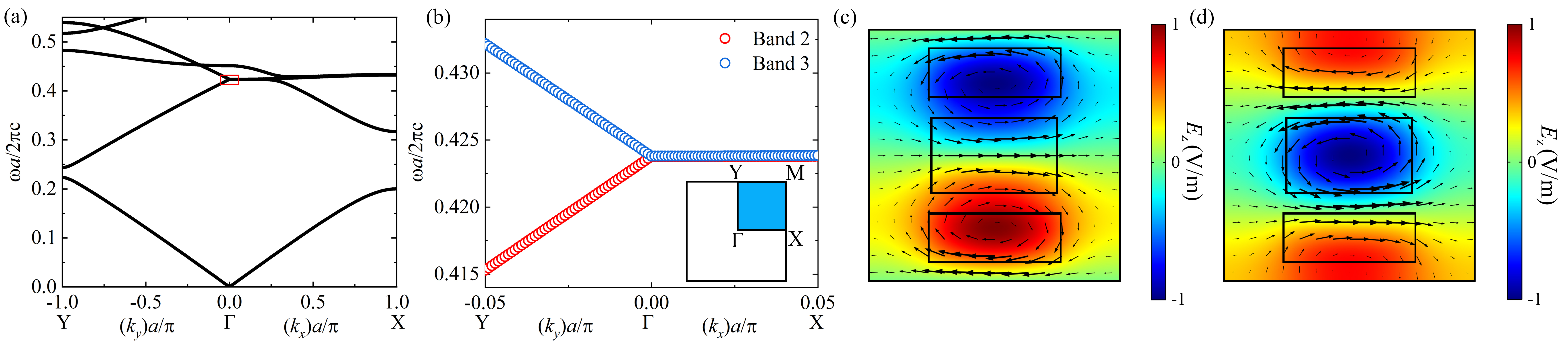}
\caption{(a) Band structures of Hermitian PC $(\gamma=0)$ computed with COMSOL. (b) The zoomed-in view of a part of the second and third bands in (a) near the $\Gamma$ point (marked by the red rectangle). (c,d) Electric field $E_z$ of bands 2 and 3, respectively. The black arrows indicate the in-plane magnetic field vector, whose magnitude is proportional to the length of the arrows.}\label{Fig2}
\end{figure*}
where $c$ is the speed of light and $\omega_{\boldsymbol{k} n}$ and $\psi$ are the eigenfrequencies and eigenvectors of the non-Hermitian system ($\gamma \neq 0$), $n$ denotes the band index and is a positive integer, $k$ denotes Bloch wave vector. $\boldsymbol{H}$ is the non-Hermitian Hamitonian matrix, and the components of $\boldsymbol{H}_1$, $\boldsymbol{H}_2$ are given by
\begin{equation}
\begin{aligned}
&\left(\boldsymbol{H}_1\right)_{m^{\prime} m}=\iint u_{\boldsymbol{k} m^{\prime}}^{(0) *}(\boldsymbol{r}) \tilde{\varepsilon}(\boldsymbol{r}) u_{\boldsymbol{k} m}^{(0)}(\boldsymbol{r}) d \boldsymbol{r}, \\
&\left(\boldsymbol{H}_2\right)_{m^{\prime} m}=\left(\frac{\omega_{m}^{(0)}}{c}\right)^2 \iint u_{\boldsymbol{k} m^{\prime}}^{(0) *}(\boldsymbol{r}) \varepsilon(\boldsymbol{r}) u_{\boldsymbol{k} m}^{(0)}(\boldsymbol{r}) d \boldsymbol{r}.
\end{aligned}\label{121}
\end{equation}
where $\omega_{\boldsymbol{k} m}^{(0)}$ and $u_{\boldsymbol{k} m}^{(0)}(\boldsymbol{r})$ are the eigenfrequency and periodic function of the $m$th band of the Hermitian system ($\gamma=0$), and $\tilde{\varepsilon}(\boldsymbol{r})=\varepsilon(\boldsymbol{r})+i \varepsilon_i(\boldsymbol{r})$ is the location-dependent complex permittivity. The periodic function $u_{\boldsymbol{k}  m}^{(0)}(\boldsymbol{r})$ and the corresponding eigenfrequency $\omega_{\boldsymbol{k} m}^{(0)}$ can be obtained using COMSOL numerical simulation. Then these periodic function $u_{\boldsymbol{k} m}^{(0)}(\boldsymbol{r})$ can be organized to satisfy
orthonormality
\begin{equation}
\int u_{\boldsymbol{k} m^{\prime}}^{(0) *}(\boldsymbol{r}) \varepsilon(\boldsymbol{r}) u_{\boldsymbol{k} m}^{(0)}(\boldsymbol{r}) \mathrm{d}^2 \boldsymbol{r}=\delta_{m^{\prime} m}\label{3},
\end{equation}
where $\delta_{m^{\prime} m}$ is the Kronecker delta function. In addition, we introduce a quantity $F_{\Omega, m^{\prime} m}$ to characterize the eigenmode profiles
\begin{equation}
F_{\Omega, m^{\prime} m}=\int_{\Omega} u_{\boldsymbol{k} m^{\prime}}^{(0) *}(\boldsymbol{r}) u_{\boldsymbol{k} m}^{(0)}(\boldsymbol{r}) \mathrm{d}^2 \boldsymbol{r}\label{4},
\end{equation}
where $\Omega$ corresponds to different medium domains.
$F_{\Omega, m m}$ expresses the amplitude distribution of the eigenmode $m$ in the domain $\Omega$, and $F_{\Omega, m^{\prime} m}\left(m \neq m^{\prime}\right)$ represents the overlapping between two different eigenmodes in the domain $\Omega$. Substituting  Eq.~(\ref{3})-(\ref{4}) into Eq.~(\ref{121}), we obtain
\begin{equation}
\left(\boldsymbol{H}_1\right)_{m^{\prime} m}=\delta_{m^{\prime} m}+i \gamma \tau_{m^{\prime} m}, \quad\left(\boldsymbol{H}_2\right)_{m^{\prime} m}=\delta_{m^{\prime} m}\left(\frac{\omega_{\boldsymbol{k} m}^{(0)}}{c}\right)^2.\label{5}
\end{equation}
It can be seen that the imaginary part of the non-Hermitian Hamiltonian $\boldsymbol{H}$ comes from the component $\boldsymbol{H}_1$, in Eq.~(\ref{5}),
\begin{equation}
\tau_{m^{\prime} m}=F_{A 1, m^{\prime} m}+F_{A 2, m^{\prime} m}+\ell_r F_{B, m^{\prime} m}\label{6}.
\end{equation}
Here, $\tau_m\left(m=m^{\prime}\right)$ is defined as the non-Hermitian strength of the $m$ state, and $\tau_{m^{\prime} m}\left(m \neq m^{\prime}\right)$ represents the coupling between two eigenmodes. Note that the integration of the eigenmode of region C without non-Hermitian perturbations is not involved in Eq.~(\ref{6}). However, region C should be considered in the computation of the orthonormalized eigenmodes. Substituting  Eq.~(\ref{4}) into Eq.~(\ref{3}), we obtain
\begin{equation}
\varepsilon_{A 1} F_{A 1, m^{\prime} m}+\varepsilon_{A 2} F_{A 2, m^{\prime} m}+\varepsilon_C F_{C, m^{\prime} m}+\varepsilon_B F_{B, m^{\prime} m}=\delta_{m^{\prime} m}\label{7}.
\end{equation}
According to the Eq.~(\ref{1})-(\ref{7}), the formulation of the non-Hermitian Hamiltonian model only needs the eigenfrequencies and the eigenfields of the Hermitian system $(\gamma=0)$.

We show the band structures of Hermitian PC in \textcolor{blue}{Fig.~\ref{Fig2}}{(a)}. The two dispersion bands (second and thirds bands) form a semi-Dirac point at the $\Gamma$ point, which is linear along the $\Gamma$Y direction but involves a quadratic band along with a flat band for the $\Gamma$X direction. \textcolor{blue}{Fig.~\ref{Fig2}}{(b)} is a zoomed-in view of the semi-Dirac dispersion near the $\Gamma$ point (marked by a red rectangle in \textcolor{blue}{Fig.~\ref{Fig2}}{(a)}). Because the second dispersion term of the second band near the $\Gamma$ point is small, it is partly degenerate with the flat band. The formation of the flat band results in the disappearance of the dispersion in the $\Gamma$X direction, implying that the group velocity of the system is zero. For a given Bloch wave\cite{enoch2003dispersion}, the averaged velocity $\mathbf{V}_e$ of the energy flow (the average is taken upon a lattice cell) is equal to the group velocity $\mathbf{V}_g$
\begin{equation}
\mathbf{V}_e=\mathbf{V}_g=\operatorname{grad}_{\mathbf{k}}(\omega)=\frac{\partial \omega}{\partial k_x} \mathbf{e}_x+\frac{\partial \omega}{\partial k_y} \mathbf{e}_y,\label{8}
\end{equation}
\noindent
\begin{figure*}[ht]
\includegraphics[width=1\textwidth]{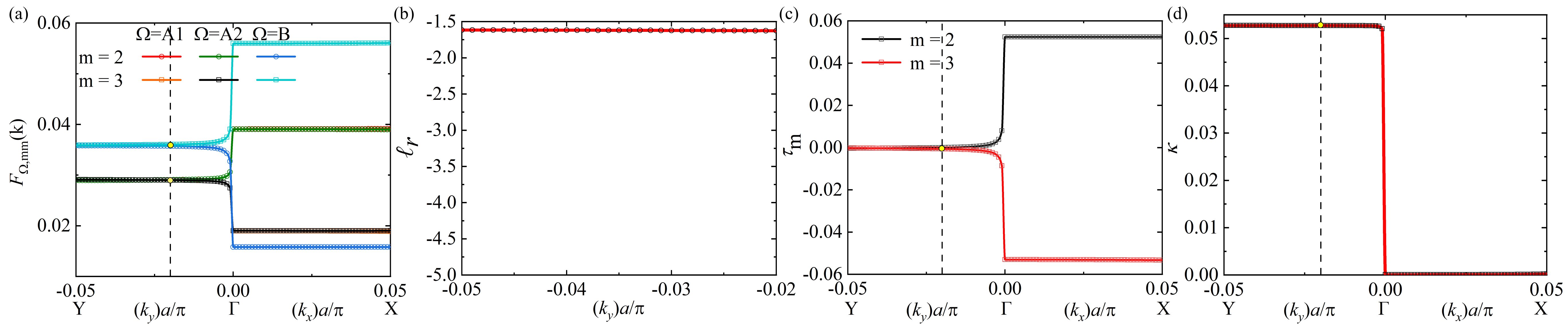}
\caption{(a) Eigenmode profiles of the second and third bands. (b) The loss-gain ratio $\ell_r$, the negative values indicate that opposite non-Hermitian perturbations are added to the red and blue regions (one region is a gain medium and the other region is a loss medium). (c) Non-Hermitian strength $\tau_m$ of the second and third bands when the loss-gain ratio $\ell_r=-1.6269$. (d) The coupling between two eigenmodes in the second and third bands. The yellow dots and vertical dashed lines correspond to $k_y=-0.02 \pi / a$ in (a),(c), and (d).}\label{Fig3}
\end{figure*}
where $\mathbf{e}_x$ and $\mathbf{e}_y$ are unit vectors. Thus, the averaged flow of energy is directly related to the dispersion curve of the Bloch mode $\mathbf{k}(\omega)$. Since the group velocity in the $k_x$ direction near the $\Gamma$ point is zero, this prevents wave propagation and presents a localized Bloch state. If the eigenmodes are examined, it is not difficult to find the underlying physical phenomena. \textcolor{blue}{Fig.~\ref{Fig2}}{(c)} and \textcolor{blue}{Fig.~\ref{Fig2}}{(d)} show the electromagnetic field distribution related to each eigenmode at the semi-Dirac point. Eigenmodes at the doubly degenerate point are composed of one dipole mode and one monopole mode. Note that the magnetic field of the dipole mode is horizontally polarized, meaning that it is a longitudinal mode along the $\Gamma$Y direction and not a transverse mode along the $\Gamma$X direction. This localized transverse magnetic field is hardly coupled to the wave with the vertically polarized magnetic field in the $\Gamma$X direction\cite{wu2014semi}. Therefore, the C2-symmetric Hermitian PC has remarkable selectivity for the direction of wave propagation. 

Considering only the contribution from the second and third bands near the semi-Dirac point, the subscript $m$ and $m^{\prime}$ in Eq.~(\ref{121})-(\ref{7}) can be represented by the numbers 2 or 3. For convenience, we can omit the subscript $k$ for simplicity and rewrite Eq.~(\ref{1}) as $\boldsymbol{H} \cdot \psi=W_n \psi$, where $W_n=\left(\omega_n / c\right)^2$. Thus, the non-Hermitian Hamiltonian model can be written as a $2 \times 2$ matrix
\begin{equation}
\boldsymbol{H}=\boldsymbol{H}_1^{-1} \cdot \boldsymbol{H}_2=\frac{1}{\beta}\left(\begin{array}{cc}
W_2^{(0)}\left(1+i \gamma \tau_2\right) & -i \gamma \kappa W_3^{(0)} \\
-i \gamma \kappa^* W_2^{(0)} & W_3^{(0)}\left(1+i \gamma \tau_3\right)
\end{array}\right),\label{9}
\end{equation}
where $\beta=1+i \gamma\left(\tau_2+\tau_3\right)+\gamma^2\left(|\kappa|^2-\tau_2 \tau_3\right)$. We use $\kappa$ instead of $\tau_{m^{\prime} m}$ to denote the overlapping coupling between the two eigenmodes, according to Eq.~(\ref{6}), 
\begin{equation}
\kappa=F_{A 1,23}+F_{A 2,23}+\ell_r F_{B, 23}.\label{10}
\end{equation}
Then the eigenvalues of the non-Hermitian Hamiltonian model can be given by
\begin{equation}
    \begin{aligned}
W_{2,3} &=\frac{1}{2 \beta}\left(\left(W_2^{(0)}+W_3^{(0)}\right) +i \gamma\left(W_2^{(0)} \tau_2+W_3^{(0)} \tau_3\right) \pm \sqrt{\Delta}\right),\\
\Delta &=-4 \gamma^2|\kappa|^2 W_2^{(0)} W_3^{(0)}\\
       &= +\left[\left(W_2^{(0)}-W_3^{(0)}\right)+i \gamma\left(W_2^{(0)} \tau_2-W_3^{(0)} \tau_3\right)\right]^2.
   \end{aligned}\label{11}
\end{equation}
In Eq.~(\ref{11}), the eigenvalues $W_{2,3}$ are real at $\tau_{2,3}=0$. To achieve the non-Hermitian strength is zero, Eq.~(\ref{6}) can be written as
\begin{equation}
F_{A 1, m m}+F_{A 2, m m}+\ell_r F_{B, m m}=0.\ (m=2,3)\label{12}
\end{equation}
So, we need a specific value of the loss-gain ratio $\ell_r$ to
satisfy Eq.~(\ref{12}), and the value of the loss-gain ratio $\ell_r$ can be solved by
\begin{equation}
\ell_r=\frac{-\left(F_{A 1, m m}+F_{A 2, m m}\right)}{F_{B, m m}}.\ (m=2,3)\label{13}
\end{equation}
To determine the value of $\ell_r$ in Eq.~(\ref{13}), the PC system must also satisfy the following conditions
\begin{equation}
F_{A 1,22}=F_{A 1,33},\ F_{A 2,22}=F_{A 2,33},\ F_{B, 22}=F_{B, 33}.\label{14}
\end{equation}
Amplitude distribution of eigenmodes in different regions (A1, A2 and B) as shown in Fig.~\ref{Fig3}{(a)}. Due to the eigenmode symmetry near the $\Gamma$ point, the eigenmode profiles of the A1 and A2 regions are similar, which behaves as $F_{\mathrm{A} 1, m m}=F_{\mathrm{A} 2, m m}$. In addition, $F_{\Omega, m m}$ does not change with the Bloch wave vectors in the $\Gamma$X direction because of the formation of the zero-dispersion bands. Note that the numerical difference between $F_{\Omega, 2 2}$ and $F_{\Omega, 3 3}$ gradually decreases along the $\Gamma$Y direction, and achieves $F_{\Omega, 22}=F_{\Omega, 33}$ when $k_y \leq-0.02 \pi / a$. Then Eq.~(\ref{13}) can be used to determine the loss-gain ratio $\ell_r$, whose value is very close to a constant (about $-1.6269$) shown in Fig.~\ref{Fig3}{(b)}. Substituting $\ell_r=-1.6269$ into Eq.~(\ref{12}), we can obtain the non-Hermitian strength of the second and third bands shown in Fig.~\ref{Fig3}{(c)}. It can be seen that the values of $\tau_m$ tend to zero when $k_y \leq-0.02 \pi / a$, which means that the non-Hermitian Hamiltonian $\boldsymbol{H}$ can have real eigenvalues. However, the values of $\tau_m$ are non-zero when $k_y>-0.02 \pi / a$, which leads the eigenvalues to have imaginary parts. Here, the yellow dot corresponds to a phase transition point in the eigenvalue spectrum, called an exceptional point (EP). According to Eq.~(\ref{11}), the EP emerges at $\Delta=0$ when the values of $\tau_m$ are zero. Near the $\Gamma$ point, the $W_{2,3}^{(0)}$ can be represented by a small $k_y$
\begin{equation}
W_{2,3}^{(0)}=\left(\frac{\omega_0}{c} \pm \frac{v_g k_y}{c}\right)^2 \approx W_0 \pm C_g k_y,\label{15}
\end{equation}
\noindent
where $C_g=2 \omega_0 v_g / c^2$, $W_0=\left(\omega_0 / c\right)^2$, and $\omega_0$ is the eigenfrequency at the semi-Dirac point, and $v_g$ is the group velocity of the linear dispersion of the Hermitian PC. Substituting Eq.~(\ref{15}) into Eq.~(\ref{11}), the positions of the EP can be given by
\begin{equation}
k_p=\left(\omega_0 / 2 v_g\right)\left(1+|\gamma \kappa|^{-2}\right)^{-\frac{1}{2}},\label{16}
\end{equation}
\begin{figure*}[ht]
\includegraphics[width=1\textwidth]{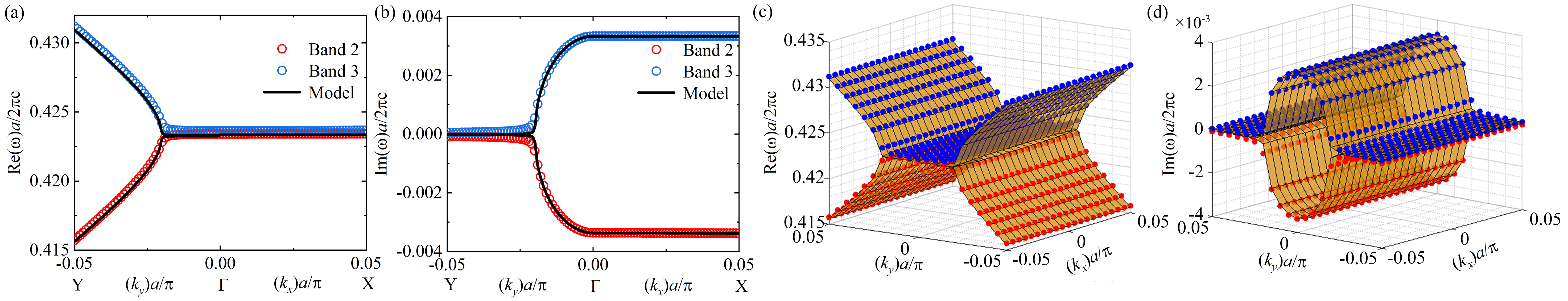}
\caption{(a,b) The real and imaginary parts of the band structures along the Y-$\Gamma$-X direction. (c,d) The real parts and imaginary parts of the band structures near the middle of the Brillouin zone. Circles and dots are calculated by COMSOL, and black lines and orange surfaces are calculated using the non-Hermitian Hamiltonian model Eq.~(\ref{11}). The parameters used are $\gamma=0.3$, $\ell_r=-1.6269$, and $\kappa=0.05272$. }\label{Fig4}
\end{figure*}
\begin{figure*}[ht]
\includegraphics[width=1\textwidth]{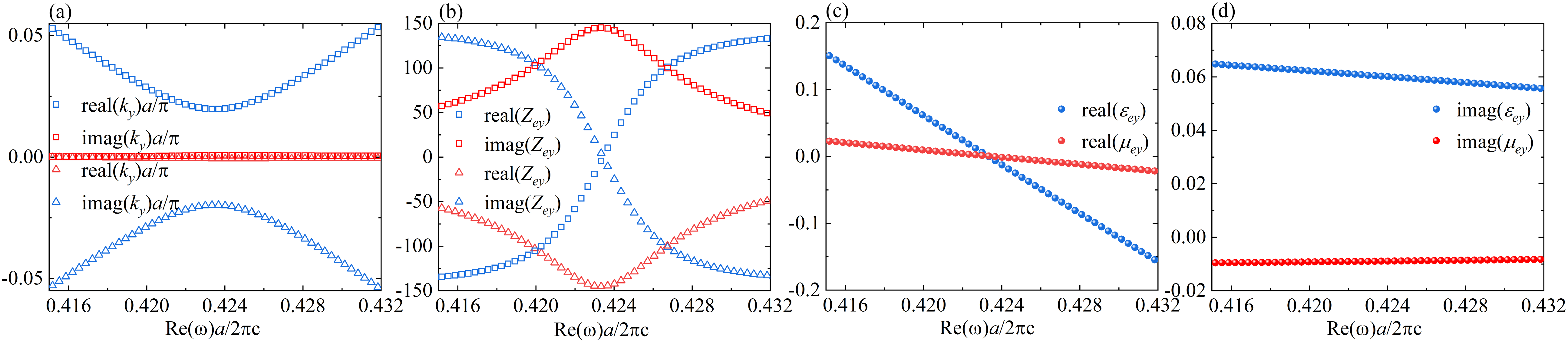}
\caption{(a) The complex Bloch $k$ bands as functions of real frequencies. (b) The real and imaginary parts of the effective impedance $Z_{e y}$. Bands with positive and negative wave vectors are represented by squares and triangles, respectively. (c,d) The real and the imaginary parts of the effective parameters $\varepsilon_{e y}$ and $\mu_{e y}$, note that effective parameters obtained from the two bands ($k_y$ and $-k_y$) are the same.}\label{Fig5}
\end{figure*}
where the value of $\kappa$ is shown in \textcolor{blue}{Fig.~\ref{Fig3}}{(d)}. It can be found that $\kappa$ is almost a constant (about 0.05272) in the $\Gamma$Y direction. According to Eq.~(\ref{16}), the non-Hermitian perturbation $\gamma$ can change the value of $k_p$. EP separates the unbroken $\Delta>0$ and broken $\Delta<0$ phases. When the condition of $\tau_m=0$ is satisfied, the upper bound of $k_p$ is $-0.02 \pi / a$, and the corresponding $\gamma=0.3$. In addition, $\kappa$ is always zero in the $\Gamma$X direction, which means the decoupling of the two eigenmodes at the semi-Dirac point. Combined with \textcolor{blue}{Fig.~\ref{Fig2}}{(b)}, the linear term disappears near the $\Gamma$ point as the coupling integral $\kappa$ disappears\cite{li2013selection}. It can be deduced that the disappearance of the linear term also has an important impact on the formation of EP. For the $\Gamma$X direction, considering that the eigenfrequency of the flat bands is $\omega_0$ in \textcolor{blue}{Fig.~\ref{Fig2}}{(b)}, and $\tau_2=-\tau_3$ in \textcolor{blue}{Fig.~\ref{Fig3}}{(c)}. The eigenvalues of Eq.~(\ref{11}) can be written as
\begin{equation}
W_m=\frac{W_0}{\left(1+i \gamma \tau_m\right)}(m=2,3)\label{17}.
\end{equation}
Eq.~(\ref{17}) shows that the eigenvalues will form a complex conjugate pair, which leads to the absence of the real spectrum of non-Hermitian PC in the $\Gamma$X direction. Now, we can verify the accuracy of the non-Hermitian model. As shown in \textcolor{blue}{Fig.~\ref{Fig4}}{(a)} and \ref{Fig4}{(b)}, we show the band structures along the Y-$\Gamma$-X direction. Black lines show the theoretical results, and circles show the simulation results. The bands calculated using the non-Hermitian Hamiltonian model Eq.~(\ref{11}) (lines) agree well with the result of simulations (circles). Additionally, we show the shape of these EPs in Bloch k-space in \textcolor{blue}{Fig.~\ref{Fig4}}{(c)} and \ref{Fig4}{(d)}. The orange surfaces (theoretical results) and dots (simulation results) indicate that these EPs form two symmetrical straight lines near the center of the Brillouin zone. The eigenfrequencies inside the EPs line form complex-conjugate pairs, and outside the EPs line, the system can obtain the real spectra. Hence, our non-Hermitian PC will exhibit different scattering behaviors for wave vectors in $\Gamma$X and $\Gamma$Y directions.

\section*{CPA and  laser in the $\Gamma$Y direction}
\begin{figure*}[!ht]
\centering
\includegraphics[width=0.9\textwidth]{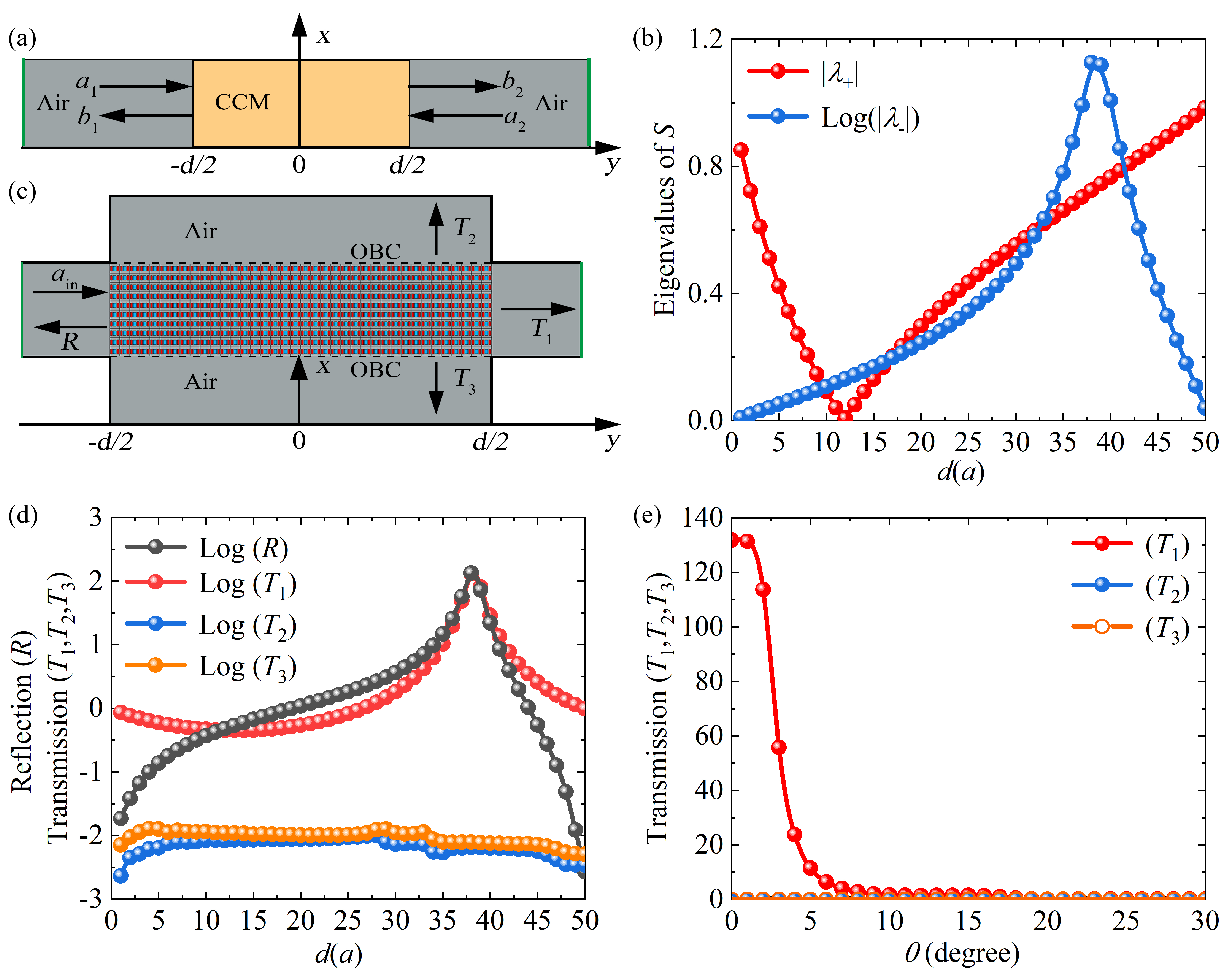}
\caption{(a) Schematic graph of the CCM, which is homogeneous and symmetric about $x = 0$, the effective parameters of CCM are shown in \textcolor{blue}{Fig.~\ref{Fig5}}. (b) Values of $\left|\lambda_{+}\right|$ (solid red dots) and $\log \left(\left|\lambda_{-}\right|\right)$ (solid blue dots) as functions of the length of the CCM slab. (c) Schematic graph of the transmission and reflection of plane waves normally incident on the width-finite PC slab. The black dashed lines represent the open boundary condition, and the solid green lines represent the ports in (a,c). (d) The reflection and transmission for the PC slab with different lengths. (e) Transmission of the PC slab with oblique incidence and the length of the PC slab consists of 38 unit cells. The plane wave with frequency $\omega=0.424(2 \pi c / a)$.}\label{Fig6}
\end{figure*}
In this section, we will describe the behavior of waves in the PC by EMT. The boundary field averaging method can calculate the effective parameters of PC. When the PC is incident by TM polarized waves along the $\Gamma$Y direction, the surface impedance of the PC is defined as
\begin{equation}
Z_{e y}=\frac{\left\langle E_z\right\rangle}{\left\langle H_x\right\rangle}=\frac{\int_{-0.5 a}^{0.5 a} E_z d x}{\int_{-0.5 a}^{0.5 a} H_x d x},\label{18}
\end{equation}
where $\left\langle E_z\right\rangle$ and $\left\langle H_x\right\rangle$ are the averaged electric and magnetic fields at the boundary of the unit cell, and $Z_{e y}=E_z / H_x=-\omega \mu_{e y} / k_y=-k_y /\left(\omega \varepsilon_{e y}\right)$. The effective permittivity and permeability can be described as follows
\begin{equation}
\varepsilon_{e y}=\frac{-k_y}{\omega \varepsilon_0 Z_{e y}}, \mu_{e y}=\frac{-k_y}{\mu_0 \omega} Z_{e y}.\label{19}
\end{equation}
Note that the effective parameters should be calculated as a function of the real frequency because the frequency of the incident wave in the actual experiment is a real number. We should use the eigenstates with complex Bloch $k$ but real-valued frequencies to retrieve the effective parameters. Here, the eigenfields $E_z$ and $H_x$ are obtained by solving the complex Bloch $k$ bands. We compute the complex Bloch $k$ bands and the corresponding eigenstates via the weak-form PDE module of COMSOL Multiphysics software package (see details in Appendix B). The obtained band dispersion is shown in \textcolor{blue}{Fig.~\ref{Fig5}}{(a)}, we can see that there is a $k_y$ gap  within the range $\left|k_y\right| a<0.02 \pi$ along the $\Gamma$Y direction, which corresponds to the broken phase in \textcolor{blue}{Fig.~\ref{Fig4}}{(a)} and \ref{Fig4}{(b)}. Furthermore, the imaginary parts of the wave vector $k_y$ are almost zero, and the effective refractive index can be given by
\begin{equation}
n_{e y}^2=\varepsilon_{e y} \mu_{e y}=\left(k_y / k_0\right)^2,\label{20}
\end{equation}
where $k_0$ is the wave vector in air. Therefore, the $n_{e y}$ of the non-Hermitian PC is a real number. \textcolor{blue}{Fig.~\ref{Fig5}}{(b)} shows the surface impedance corresponding to positive and negative wave vectors. Then, according to Eq.~(\ref{19}), we can calculate the effective parameter of the non-Hermitian PC. As shown in \textcolor{blue}{Fig.~\ref{Fig5}}{(c)} and \ref{Fig5}{(d)}, real parts of the effective parameters $\varepsilon_{e y}$ and $\mu_{e y}$ approach zero simultaneously at the EP frequency ($\omega=0.424(2 \pi c / a)$), while both possess non-zero imaginary parts. Because the $\varepsilon_{e y}$ and $\mu_{e y}$ are purely imaginary numbers, this non-Hermitian PC behaves like the homogeneous CCM, which the CPA or laser can be achieved at EP frequency. As shown in \textcolor{blue}{Fig.~\ref{Fig6}}{(a)}, We consider two counterpropagating plane waves with the same frequency $\omega$ incident on a CCM slab (orange region) with thickness $d$ embedded in air. The incident plane waves are $E_z$ polarized, and the wave vector is along the $y$ direction. Then, the electric fields in the background (air) can be  expressed as\cite{bai2016simultaneous}
\begin{equation}
E_z=\left\{\begin{array}{l}
a_1 e^{i k_0 y}+b_1 e^{-i k_0 y}\quad y \leqslant-d / 2, \\
b_2 e^{i k_0 y}+a_2 e^{-i k_0 y}\quad y \geqslant-d / 2,
\end{array}\right.\label{21}
\end{equation}
where $k_0=\omega / c_0$, $a_i, b_i, i=1,2$ are the coefficients of the right and left propagating waves. In terms of the
scattering matrix $S$, we can write
\begin{equation}
\begin{aligned}
\left(\begin{array}{l}
b_2 \\
b_1
\end{array}\right) &=S\left(\begin{array}{l}
a_1 \\
a_2
\end{array}\right)=\left(\begin{array}{ll}
T & R \\
R & T
\end{array}\right)\left(\begin{array}{l}
a_1 \\
a_2
\end{array}\right) \\
&=\frac{1}{M_{22}}\left(\begin{array}{cc}
1 & M_{12} \\
M_{12} & 1
\end{array}\right)\left(\begin{array}{l}
a_1 \\
a_2
\end{array}\right),
\end{aligned}\label{22}
\end{equation}
where $T$ and $R$ represent the transmission and reflection 
coefficients of the CCM slab, $T=1 / M_{22}$, and $R=M_{12} / M_{22}$. The elements in $S$ are
\begin{equation}
M_{12}=i \frac{n_{e y}^2-\mu_{e y}^2}{2 n_{e y} \mu_{e y}} \sin (\eta),\label{23}
\end{equation}
\begin{equation}
M_{22}=e^{i d k_0}\left[\cos (\eta)-i \frac{n_{e y}{ }^2+\mu_{e y}^2}{2 n_{e y} \mu_{e y}} \sin (\eta)\right],\label{24}
\end{equation}
where $\eta=n_{e y} k_0 d$, the eigenvalues and eigenvectors of the scattering matrix are $\lambda_{\pm}=T \pm R=\left(1 \pm M_{12}\right) / M_{22}$ and $\vec{\varphi}_{\pm}=(1, \pm 1)^T$.
Therefore, the outgoing waves can be obtained
\begin{equation}
\left(\begin{array}{l}
b_2 \\
b_1
\end{array}\right)=\frac{1}{2}\left[\left(a_1+a_2\right) \lambda_{+} \vec{\varphi}_{+}+\left(a_1-a_2\right) \lambda_{-} \vec{\varphi}_{-}\right].\label{25}
\end{equation}
At $\omega=0.424(2 \pi c / a)$, the variation of the absolute values of the eigenvalues $\left|\lambda_{\pm}\right|$ with the length $d(a)$
shown in \textcolor{blue}{Fig.~\ref{Fig6}}{(b)}. It can be seen that $\left|\lambda_{+}\right|$ is close to zero when $d=12a$ and $\left|\lambda_{-}\right|$ reaches the maximum value at $d=38a$. Therefore, CPA can be achieved by setting $a_1=a_2$ (in-phase excitation), and the CCM slab has 12 layers, while the laser can be realized by setting $a_1=-a_2$ (out-of-phase excitation) and the CCM slab has 38 layers. \textcolor{blue}{Fig.~\ref{Fig6}}{(c)} shows the transmission and reflection of a plane wave at EP frequency along the $y$-direction. The PC slab only contains 10 unit cells in the $x$ direction, while it can have a different number of unit cells in the $y$ direction. Note that we use open boundary condition (OBC) to simulate the leaky wave along the $x$ direction, which does not add additional confining metallic materials to constrain the propagation of the light. As shown in \textcolor{blue}{Fig.~\ref{Fig6}}{(d)}, it can be seen that the transmission $T_1$ is at least 100 times higher than the transmission  $T_2$ and $T_3$, which indicates that the leaky wave in the $x$ direction is feeble. Furthermore, the values of reflection and transmission are almost the same at $d=12a$. Substituting $\left|\lambda_{+}\right|=0$ and $a_2=0$ into Eq.~(\ref{25}), we can obtain $R=T_1=\left|\lambda_{-}\right| / 2$. Then, the values of reflection and transmission will peak because of the divergence of $\left|\lambda_{-}\right|$ at $d=38a$. Because the transmission efficiency of the incident wave varies greatly in different directions, the PC slab can be used as a sensor with angle selection functionality. As shown in \textcolor{blue}{Fig.~\ref{Fig6}}{(e)}, it clearly shows that at normal incidence, the transmission is as high as 130 and then drops to $10 \%$ of the peak value when the angle of incidence is only $5^{\circ}$. $T_2$ and $T_3$ are near zero even though the angle of incident wave changes continuously, meaning that the PC prevents the spread of wave in the $x$ direction.
\begin{figure}[!htbp]
\includegraphics[width=0.45\textwidth]{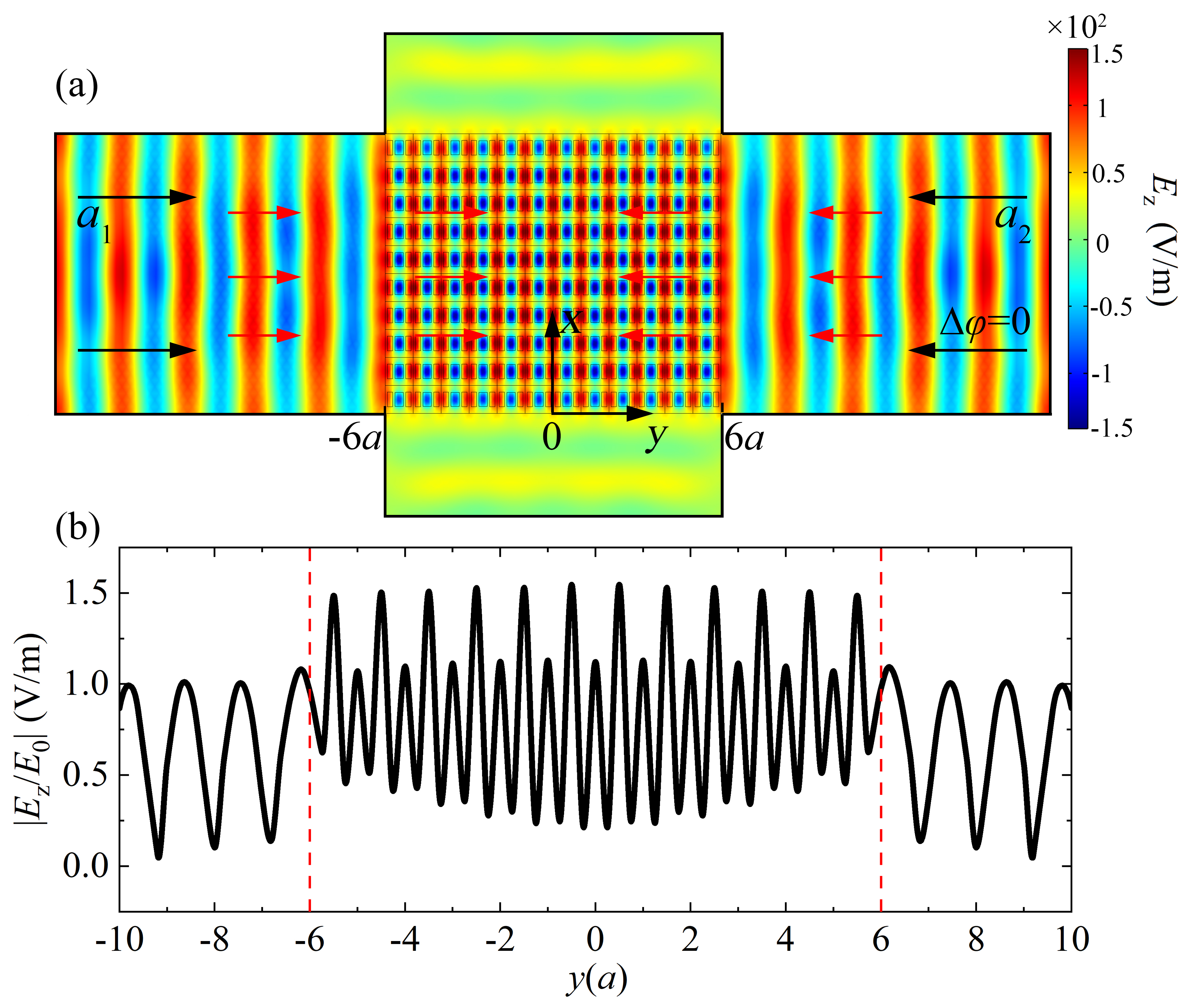}
\caption{CPA. (a) The field distributions of the electric field $E_z$ in the PC slab. Poynting vectors (red arrows) for two counterpropagating plane waves (black arrows) normally incident
on the PC slab in the case of $\Delta \varphi=0$, and the PC slab consists of 12 layers of the unit cell. (b) The normalized amplitude of the electric field $\left|E_z / E_0\right|$ for the case of (a), and the red dashed lines denote the boundaries of the PC slab. The two plane waves with frequency $\omega=0.424(2 \pi c / a)$.}\label{Fig7}
\end{figure}
\begin{figure}[!htbp]
\includegraphics[width=0.45\textwidth]{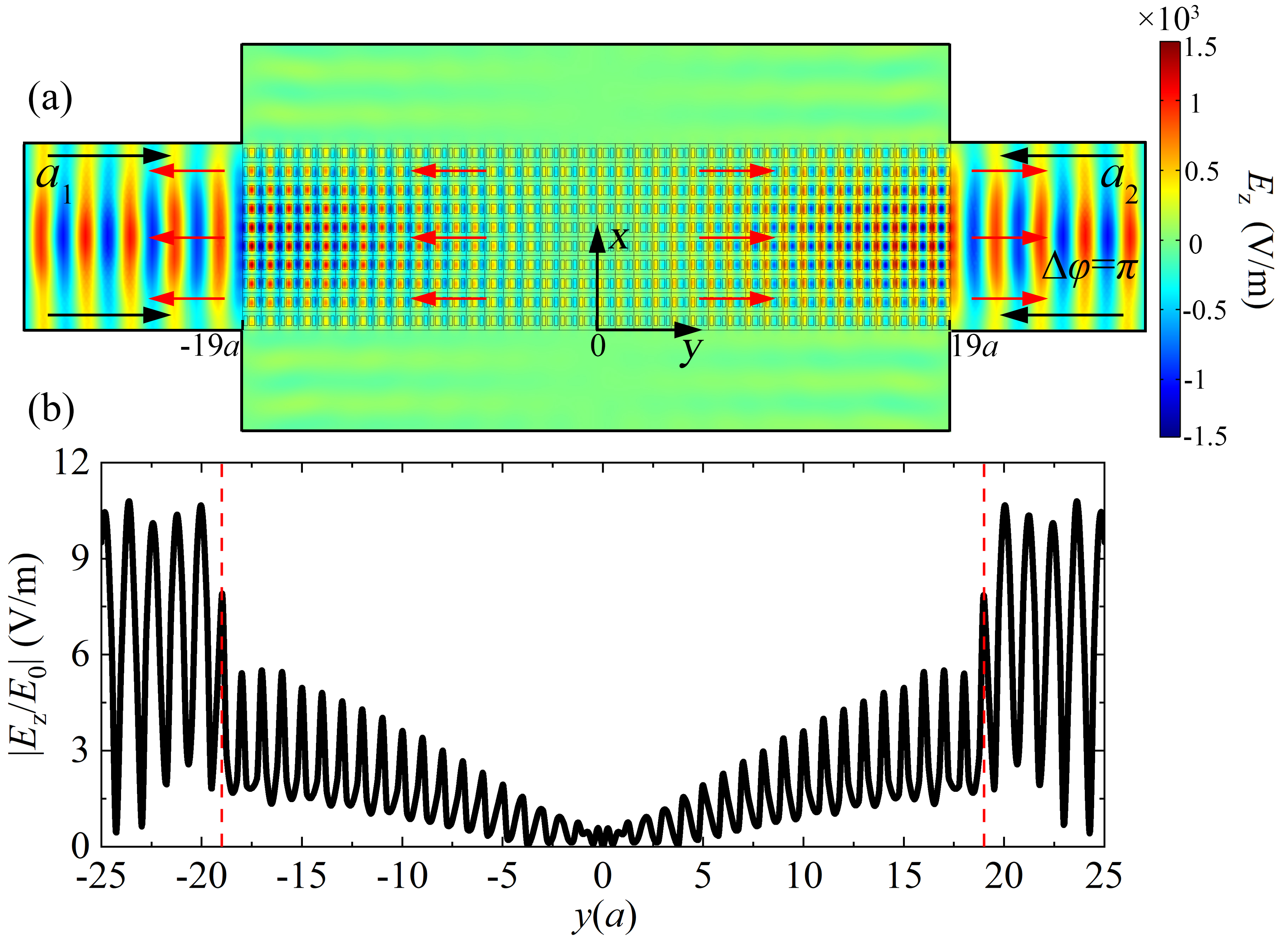}
\caption{Laser. (a) The field distributions of electric field $E_z$ in the PC slab. Poynting vectors (red arrows) for two counterpropagating plane waves (black arrows) normally incident
on the PC slab in the case of $\Delta \varphi=\pi$, and the PC slab consists of 38 layers of the unit cell. (b) The normalized amplitude of the electric field $\left|E_z / E_0\right|$ for the case of (a), and the red dashed lines denote the boundaries of the PC slab. The two plane waves with frequency $\omega=0.424(2 \pi c / a)$.}\label{Fig8}
\end{figure}

The effect of the CPA  by using the PC slab is shown in \textcolor{blue}{Fig.~\ref{Fig7}}{(a)}, which shows the field distribution of the electromagnetic wave $E_z$ when two coherent counter-propagating waves are normally incident on the PC slab with a phase difference $\Delta \varphi=0$. The Poynting vectors shown by the red arrows point inward, which indicates that the incident waves are absorbed by the PC slab. Note that the electric field outside the OBC is close to zero, and the corresponding leaky wave in the $x$ direction can be ignored. Besides, we plot the amplitudes (normalized to the incident wave) of the electric field shown in \textcolor{blue}{Fig.~\ref{Fig7}}{(b)}. Obviously, for the CPA case, $\left|E_z\right|$ outside the PC slab approaches $\left|E_0\right|$, meaning that the reflection at the boundary is tiny and almost all the incident energy is absorbed. On the other hand, in the case of $\Delta\varphi=\pi$, the power flows are entirely reversed. As shown in \textcolor{blue}{Fig.~\ref{Fig8}}{(a)}, the Poynting vectors are now all pointing outwards, which indicates the energy transfer from the PC slab to the air background. The normalized amplitude of the electric field in the background is shown in \textcolor{blue}{Fig.~\ref{Fig8}}{(b)}. It can be seen that the fields outside the PC slab are significantly amplified, which are about ten times the incident waves, indicating a laser phenomenon. However, the amplitude of the field outside the OBC is still extremely low, meaning that the waves are not enhanced in the $x$ direction. It can be deduced that the waves will exhibit independent propagation behavior in different directions, corresponding to an anisotropic non-Hermitian PC. In the following subsection, we will show the formation of evanescent in the $x$ direction.

\section*{Formation of evanescent waves for the $\Gamma$X direction}
We have shown in the previous subsection that the system can have are real spectrum in the $\Gamma$Y direction, and the PC behaves as a CCM with a real refractive index, which can realize CPA and laser. However, the non-Hermitian Hamiltonian model shows that the eigenvalues have imaginary parts for the $\Gamma$X direction, implying that the effective refractive index will be complex. 
\begin{figure}[!htbp]
\centering
\includegraphics[width=0.45\textwidth]{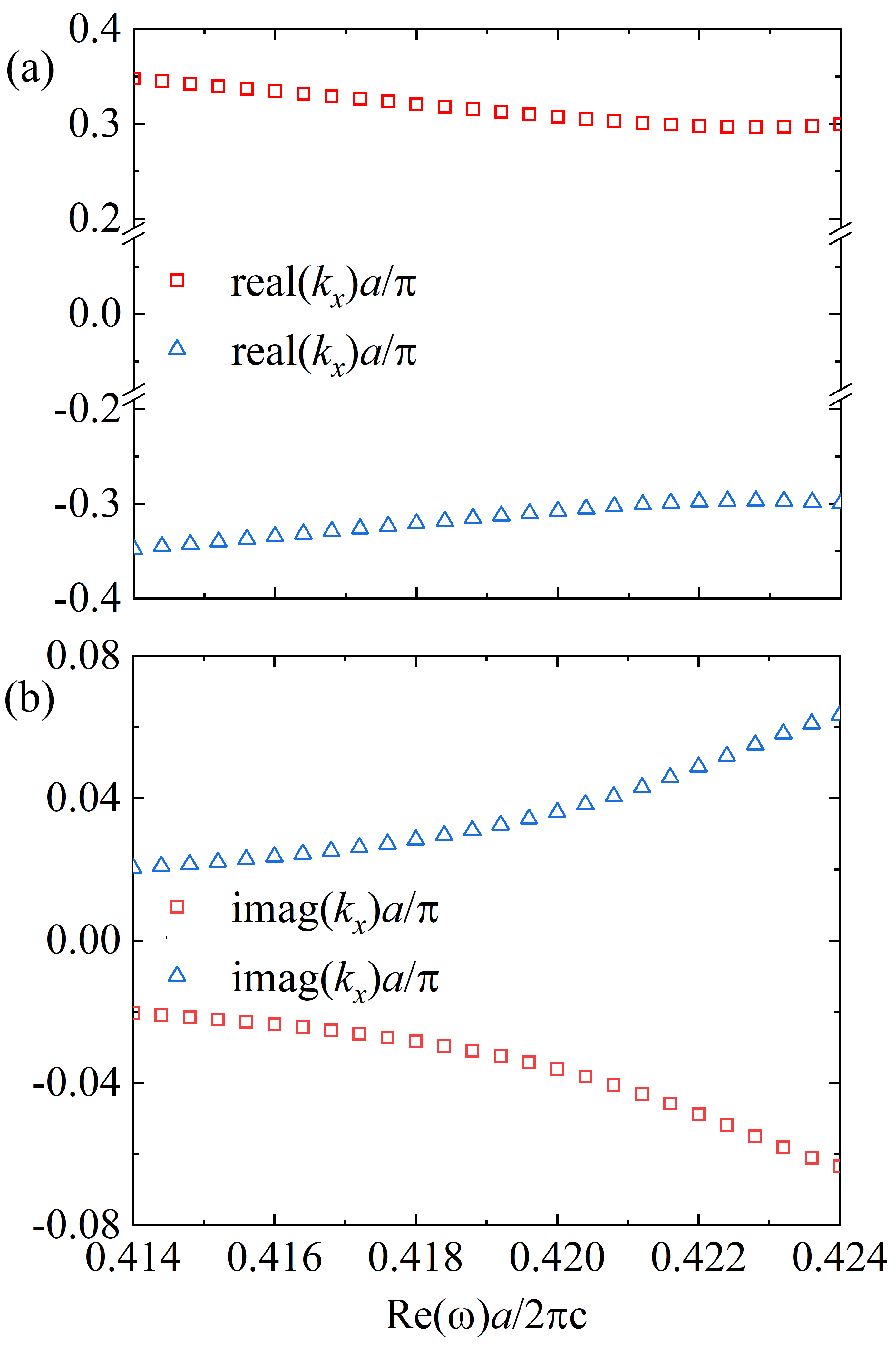}
\caption{(a) The real parts of the complex Bloch $k$ bands as functions of real frequencies. (b) The imaginary parts of the complex Bloch $k$ bands as functions of real frequencies. Bands with positive and negative wave vectors are represented by squares and triangles, respectively.}\label{Fig9}
\end{figure}
In the actual experiments, the frequencies of the incident wave are real, so we need to establish the EMT to study the scattering properties of the wave in the $\Gamma$X direction. The effective parameters for the $\Gamma$X direction can be expressed as
\begin{equation}
\varepsilon_{e x}=\frac{-k_x}{\omega \varepsilon_0 Z_{e x}}, \mu_{e x}=\frac{-k_x}{\mu_0 \omega} Z_{e x},\label{26}
\end{equation}
where the surface impedance $Z_{e x}=\left\langle\mathrm{E}_z\right\rangle /\left\langle H_y\right\rangle$, and $\left\langle E_z\right\rangle$ and $\left\langle H_y\right\rangle$ are the averaged electric and magnetic fields at the boundary of the unit cell, which can be obtained by solving the complex Bloch $k$ bands. \textcolor{blue}{Fig.~\ref{Fig9}}{(a)} shows the real part of the Bloch $k$ bands calculated by the weak-form PDE module in COMSOL,
\begin{figure}[!htbp]	
\centering
\includegraphics[width=0.45\textwidth]{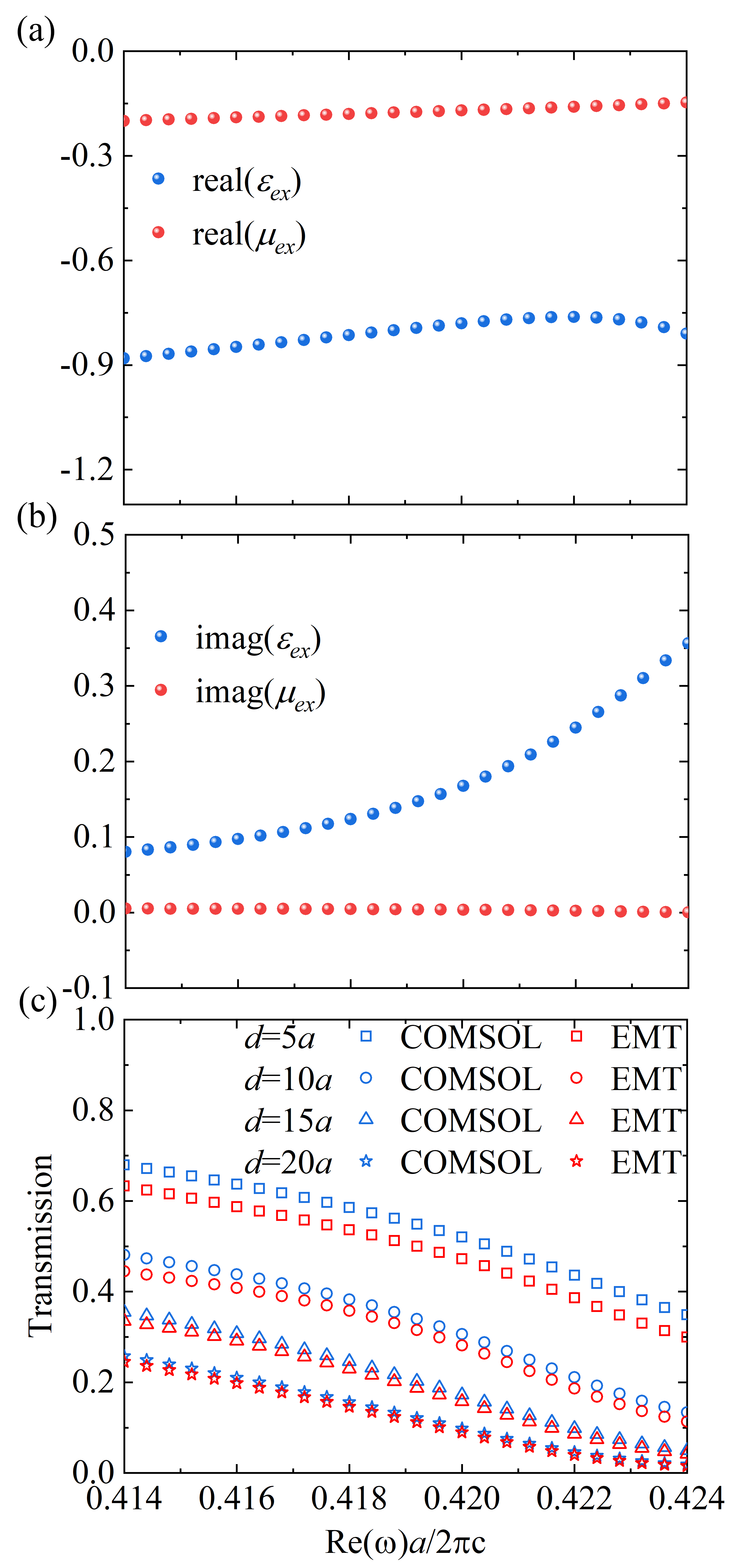}
\caption{(a,b) The real and the imaginary parts of the effective parameters $\varepsilon_{e x}$ and $\mu_{e x}$, note that effective parameters obtained from the two bands ($k_x$ and $-k_x$) are the same. (c) The transmission spectrums of PC with different lengths are calculated using COMSOL simulation and EMT, respectively. The incident plane waves are $E_z$ polarized and come from the $\Gamma$X direction.}\label{Fig10}
\end{figure}
where the triangles and squares represent the positive and negative wave vectors, respectively. Note that the Bloch $k$ bands correspond to a part of the quadratic band (Band 2 shown in \textcolor{blue}{Fig.~\ref{Fig2}}{(a)}) in the $\Gamma$X direction, and the difference is due to the ignoring of the flat bands in the broken phase. Only the quadratic band has a large group velocity to support wave propagation in PC, and its eigenfields can be used to extract the effective parameters. Besides, we show the imaginary part of the Bloch $k$ bands in \textcolor{blue}{Fig.~\ref{Fig9}}{(b)}. Combined with \textcolor{blue}{Fig.~\ref{Fig9}}{(a)}, it can be seen that the wave vector $k_x$ in the positive or negative directions, the signs of its imaginary parts are opposite. For the non-Hermitian PC systems, the eigenfields can be expressed as
\begin{equation}
E_{\boldsymbol{k_x} n}(\boldsymbol{r})=u_{\boldsymbol{k_x} n}(\boldsymbol{r}) e^{-i \boldsymbol{k_x} \cdot \boldsymbol{r}},\label{27}
\end{equation}
among them, $ u_{\boldsymbol{k_x} n}$ is the periodic function of the $n$th band, and $n$ denotes the Bloch $k_x$ band index. Here, the wave vector $k_x$ is a complex number, considering that the real and imaginary parts of wave vector $k_x$ have opposite signs. Thence, the Bloch state in Eq.~(\ref{27}) can be rewritten as
\begin{equation}
E_{\boldsymbol{k_x} n}(\boldsymbol{r})=u_{\boldsymbol{k_x} n}(\boldsymbol{r}) e^{-i \cdot \operatorname {r e a l}\left(\boldsymbol{k_x}\right) \cdot \boldsymbol{r}} \cdot e^{\operatorname {i m a g}\left(\boldsymbol{k_x}\right) \cdot \boldsymbol{r}},\label{28}
\end{equation}
where $e^{\operatorname{imag}\left(\boldsymbol{k_x}\right) \cdot \boldsymbol{r}}$ is a decay factor, indicating that the amplitude of the eigenfields $E_z$  will decrease with the propagation distance. It can be found that when the frequency $\omega$  approaches the EP frequency ($\omega=0.424(2 \pi c / a)$), the absolute value of the imaginary part of $k_x$ becomes larger. Especially at EP frequency, the eigenfields will be strongly attenuated, and an evanescent wave will form in the PC due to the effect of the decay factor. On the other hand, according to $n_{e x}^2=\varepsilon_{e x} \mu_{e x}=\left(k_x / k_0\right)^2$, the effective refractive index $n_{e x}$ is complex. The imaginary parts of $n_{e x}$ and the decay factor $e^{\operatorname{imag}\left(\boldsymbol{k_x}x\right) \cdot \boldsymbol{r}}$ are derived from the imaginary part of the Bloch $k$. 

\textcolor{blue}{Fig.~\ref{Fig10}}{(a)} and \textcolor{blue}{Fig.~\ref{Fig10}}{(b)} shows the effective parameters $\varepsilon_{e x}$ and $\mu_{e x}$. It clearly shows that the real parts of $\varepsilon_{e x}$ and $\mu_{e x}$ do not pass through zero. The imaginary parts of $\mu_{e x}$ are near zero, and only the imaginary parts of $\varepsilon_{e x}$ are non-zero. Therefore, at the EP frequency, the imaginary part of $n_{e x}$ has a maximum value because the imaginary parts of $\varepsilon_{e x}$ reach a peak. To analyze the scattering properties of PC, we assume that a wave is normally incident along the $\Gamma$X direction to the PC with different lengths $d$. Substituting the effective parameters $\varepsilon_{e x}$ and $\mu_{e x}$ into Eq.~(\ref{24}), the transmission spectrum can be calculated by EMT. As shown in \textcolor{blue}{Fig.~\ref{Fig10}}{(c)}, we give the results of the COMSOL simulation and EMT. The results are well consistent, confirming the accuracy of the effective parameters shown in \textcolor{blue}{Fig.~\ref{Fig10}}{(a)} and \textcolor{blue}{Fig.~\ref{Fig10}}{(b)}. The transmission decreases with the increase of length $d$, indicating that the wave attenuates continuously with the propagation distance. When the PC length $d$ is constant, the transmittance is related to the imaginary part of $n_{e x}$, whose values determine the attenuation strength of the electric field. At the EP frequency, the transmission reaches a minimum, corresponding to the largest value of the imaginary part of $n_{e x}$. However, away from the EP frequency, the transmission gradually increases due to the decrease of the imaginary part of $n_{e x}$. 

\begin{figure}[!htbp]
\centering
\includegraphics[width=0.45\textwidth]{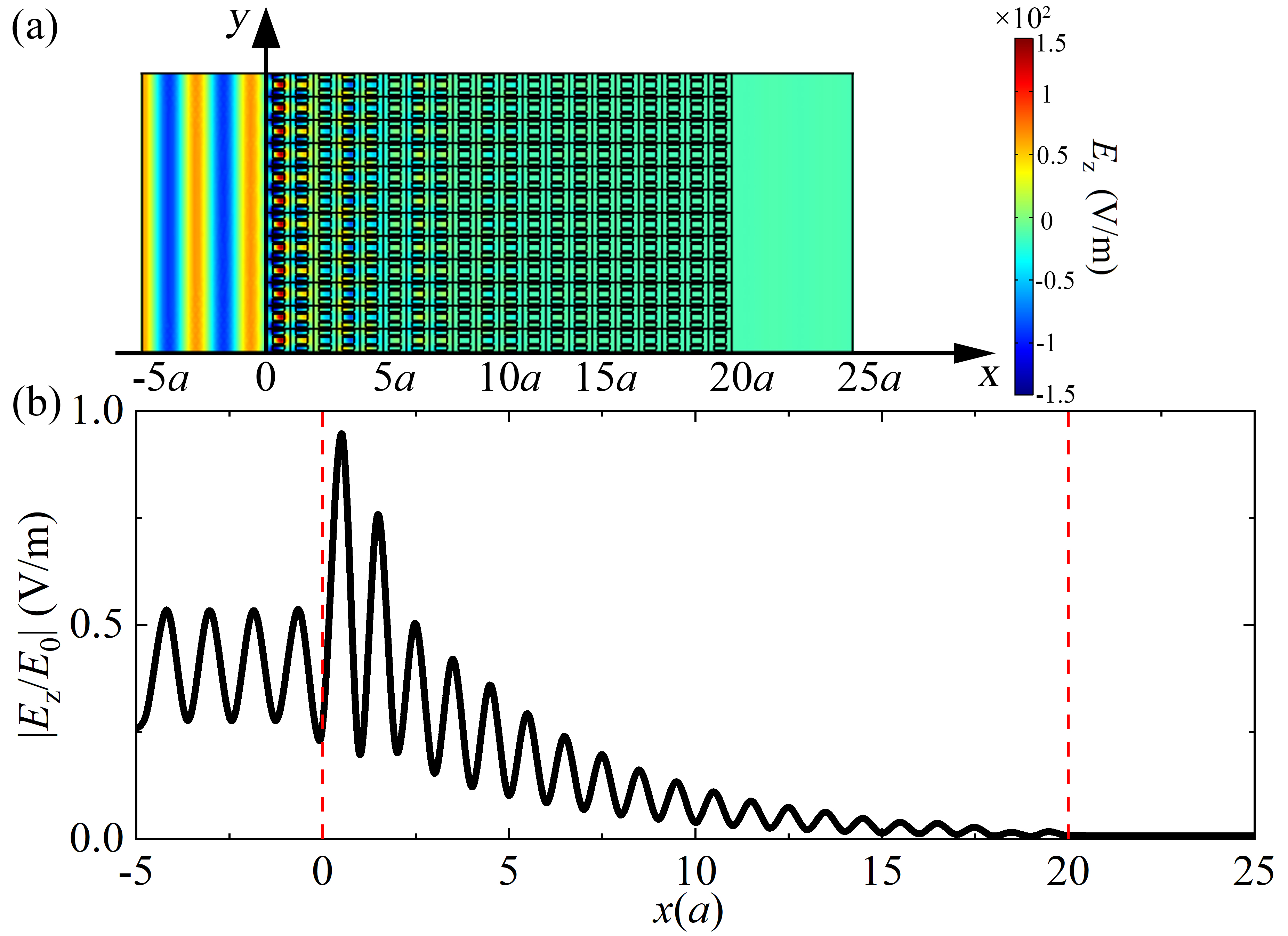}
\caption{(a) The field distributions of the electric field $E_z$ in the PC slab. (b) The normalized amplitude of electric field $\left|E_z / E_0\right|$ for the case of (a). The incident plane waves are $E_z$ polarized with
frequency $\omega=0.424(2 \pi c / a)$, and the wave vector is along the $x$ direction. As for the $y$ direction, we use the periodic boundaries to simulate an infinite PC.}\label{Fig11}
\end{figure}

In \textcolor{blue}{Fig.~\ref{Fig11}}{(a)}, we show that a plane wave with frequency $0.424(2 \pi c / a)$ is normally incident on the PC, which consists of 20 layers of the unit cell. It can be found that the wave decays rapidly in the PC, and the amplitude of the electric field of the outgoing wave is close to zero. \textcolor{blue}{Fig.~\ref{Fig11}}{(b)} shows the electric field (normalized to the incident wave) in the PC. Due to the lack of transverse modes that support wave propagation in the $x$ direction, there are some reflected waves at the boundary, which leads to $E_z$ being smaller than $E_0$ when $x<0$. In addition, the weakening of the electric field is noticeable. The amplitude of the electric field is almost zero when $x>10a$. The above results confirm the formation of evanescent waves along the $\Gamma$X direction. 

\section*{\label{sec:level4}Conclusion}
In summary, we implement a non-Hermitian PC with anisotropic emission by introducing gains and losses. The effective Hamiltonian model shows that the EP point and real spectrum appear in the $\Gamma$Y direction because of the near-zero non-Hermitian intensity $\tau_m$. However, for the $\Gamma$X direction, a non-zero $\tau_m$ leads to the absence of the EP point and real spectrum. For the $\Gamma$Y direction, CMT indicates that the PC behaves as a CCM with the real refractive index, supporting the CPA and Laser effects. However, for the  $\Gamma$X direction, the effective refractive index is a complex number with a large imaginary part corresponding to the formation of the evanescent waves along the $x$ direction. Combined with boundary reflection caused by the absence of transverse modes, we achieve the CPA and laser effects in the directional emission of the $\Gamma$Y. The outgoing waves in the undesirable direction are extremely weak even if the boundaries of the PC do not add any confining metallic materials. Therefore, this non-Hermitian PC with anisotropic emissions will reduce loss and electromagnetic interference caused by leaky waves. Because our PC has significant transmission differences in different directions,  it can have some typical applications, such as significant amplification of the signal and collimation and angle sensors.  

\section*{Acknowledgement}
We are supported by the Fundamental Research Funds for the Central University of China; the open fund of China Ship Development and Design Centre [XM0120190196]; Hongque Innovation Center [HQ202104001]; National Natural Science Foundation of China [NSFC42274206, NSFC41974195].

\setcounter{equation}{0}
\setcounter{figure}{0}
\setcounter{table}{0}

\renewcommand{\theequation}{A\arabic{equation}}
\renewcommand{\thetable}{A\arabic{table}}
\renewcommand{\thefigure}{A\arabic{figure}}
\appendix

\section*{Appendix A: The effective eigenequation of the non-Hermitian system}
For a Hermitian system, we focus on the TM polarization and the eigenvalue problem of a two-dimensional PC, which can be described by a Helmholtz equation as
\begin{equation}
\nabla \times\left(\frac{1}{\mu}(\boldsymbol{r}) \nabla \times \boldsymbol{E}(\boldsymbol{r})\right)-\varepsilon(\boldsymbol{r}) \frac{\omega^2}{c^2} \boldsymbol{E}(\boldsymbol{r})=0, \label{appA1}
\end{equation}
where $\omega$ is the angular frequency and $c$ is the speed of light. Therefore, the Bloch states for the Hermitian system can be expressed as $E_{\boldsymbol{k} n}^{(0)}(\boldsymbol{r})=u_{\boldsymbol{k} n}^{(0)}(\boldsymbol{r}) e^{-i \boldsymbol{k} \cdot \boldsymbol{r}}$, where $k$ is the Bloch wave vector, $u_{\boldsymbol{k} n}^{(0)}$ is a periodic function corresponding to the eigenfield distribution, $n$ represents the band index and is a positive integer. $\left\{u_{\boldsymbol{k} n}^{(0)}\right\}$ forms a complete orthogonal set due to the Hermiticity of the system. Therefore, the eigenstates of the non-Hermitian system $E_{\boldsymbol{k} n}(\boldsymbol{r})=u_{\boldsymbol{k} n}(\boldsymbol{r}) e^{-i \boldsymbol{k} \cdot \boldsymbol{r}}$ can be expanded in series of $\left\{u_{\boldsymbol{k} n}^{(0)}\right\}$ as
\begin{equation}
u_{\boldsymbol{k} n}(\boldsymbol{r})=\sum_{m=1}^{+\infty} \alpha_{n, \boldsymbol{k} m} u_{\boldsymbol{k} m}^{(0)}(\boldsymbol{r}), \label{appA2}
\end{equation}
where $\alpha_{n, \boldsymbol{k} m}$ is the expansion coefficient to be determined. Next, we consider that the relative permittivity of the medium in the non-Hermitian system becomes the complex numbers, Eqs.~(\ref{appA1}) can be rewritten as
\begin{equation}
\nabla \times\left[\frac{1}{\mu}(\boldsymbol{r}) \nabla \times \boldsymbol{E}(\boldsymbol{r})\right]-\frac{\omega^2}{c^2} \tilde{\varepsilon}(\boldsymbol{r}) \boldsymbol{E}(\boldsymbol{r})=0, \label{appA3}
\end{equation}
where $\tilde{\varepsilon}(\boldsymbol{r})=\varepsilon(\boldsymbol{r})+i \varepsilon_i(\boldsymbol{r})$ is the location-dependent complex
permittivity. Substituting Eqs.~(\ref{appA2}) into Eqs.~(\ref{appA3}), we obtain
\begin{equation}
\sum_{m=1}^{+\infty} \alpha_{n, \boldsymbol{k} m}\left(\frac{\omega_{\boldsymbol{k} m}^{(0)}}{c}\right)^2 \varepsilon(\boldsymbol{r}) u_{\boldsymbol{k} m}^{(0)}(\boldsymbol{r})=\frac{\omega_{\boldsymbol{k} n}{ }^2}{c^2} \tilde{\varepsilon}(\boldsymbol{r}) \sum_{m=1}^{+\infty} \alpha_{n, \boldsymbol{k} m} u_{\boldsymbol{k} m}^{(0)}(\boldsymbol{r}),\label{appA4}
\end{equation}
where $\omega_{\boldsymbol{k} m}^{(0)}$ and $\omega_{\boldsymbol{k} n}$ are the eigenfrequencies of the $m$th band of the Hermitian system and the non-Hermitian system, respectively. Multiplying Eqs.~(\ref{appA4}) by $u_{\boldsymbol{k} m^{\prime}}^{(0) *}(\boldsymbol{r})$ and integrating within a unit cell, we obtain the effective eigenequation
\begin{equation}
\left(\boldsymbol{H}_1\right)^{-1} \cdot \boldsymbol{H}_2 \cdot \psi=\boldsymbol{H} \cdot \psi=\left(\frac{\omega_{\boldsymbol{k} n}}{c}\right)^2 \psi, \label{appA5}
\end{equation}
where $\psi=\left(\cdots \alpha_{n, \boldsymbol{k} m}, \cdots\right)^T$ are the eigenvectors, $\boldsymbol{H}$ is the non-Hermitian Hamiltonian, and
the components of $\boldsymbol{H}_1$, $\boldsymbol{H}_2$ are given by
\begin{equation}
\begin{aligned}
&\left(\boldsymbol{H}_1\right)_{m^{\prime} m}=\iint u_{\boldsymbol{k} m^{\prime}}^{(0) *}(\boldsymbol{r}) \tilde{\varepsilon}(\boldsymbol{r}) u_{\boldsymbol{k} m}^{(0)}(\boldsymbol{r}) d \boldsymbol{r}, \\
&\left(\boldsymbol{H}_2\right)_{m^{\prime} m}=\left(\frac{\omega_{m}^{(0)}}{c}\right)^2 \iint u_{\boldsymbol{k} m^{\prime}}^{(0) *}(\boldsymbol{r}) \varepsilon(\boldsymbol{r}) u_{\boldsymbol{k} m}^{(0)}(\boldsymbol{r}) d \boldsymbol{r}.
\end{aligned}\label{appA6}
\end{equation}

\setcounter{equation}{0}
\setcounter{figure}{0}
\setcounter{table}{0}

\renewcommand{\theequation}{B\arabic{equation}}
\renewcommand{\thetable}{B\arabic{table}}
\renewcommand{\thefigure}{B\arabic{figure}}
\appendix

\section*{Appendix B: \texttt{Comsol} weak form equation}

Here, we compute the complex Bloch $k$ bands via Weak Form PDE interface of COMSOL Multiphysics software package. 
As before, the out-of-plane electric field can be expressed as
\begin{equation}
{E}(\boldsymbol{r})={u}(\boldsymbol{r}) e^{-i \boldsymbol{k} \cdot \boldsymbol{r}}.\label{appB1}
\end{equation}
Substituting Eqs.~(\ref{appB1}) into Eqs.~(\ref{appA3}), we obtain
\begin{equation}
\begin{aligned}
&\frac{\boldsymbol{k}^2}{\mu} u-\frac{\boldsymbol{k}}{\mu}(\boldsymbol{k} \cdot u)-i \boldsymbol{k} \times\left(\frac{1}{\mu} \times u\right)-i \nabla \times\left(\frac{1}{\mu} \boldsymbol{k} \times u\right)+\\
&\nabla \times\left(\frac{1}{\mu} \nabla \times u\right)-\tilde{\varepsilon} \frac{\omega^2}{c^2} u=0.\label{appB2}
\end{aligned}
\end{equation}
Multiplying the test function $\tilde{E_z}$ and integrating within a unit cell, the corresponding weak form is obtained as follows\cite{fietz2011complex}
\begin{equation}
\operatorname{weak}=(i \boldsymbol{k}+\nabla) \times \tilde{E}_z \cdot \frac{1}{\mu} \cdot(-i \boldsymbol{k}+\nabla) \times E_z-\tilde{E}_z \frac{\tilde{\mathcal{\varepsilon}} \omega^2}{c^2} E_z, \label{appB3}
\end{equation}
where the variable $u$ is denoted as $E_z$ in the weak form solution of the two-dimensional electromagnetic wave equation. The simulation domain in COMSOL is a unit cell, and the periodicity of $u$ is enforced by imposing periodic boundary conditions on the exterior boundaries of the unit cell. According to the Maxwell
equation, the magnetic fields are calculated by
\begin{equation}
H=\frac{1}{i \omega \mu \mu_0} \cdot\left[i \boldsymbol{k} \times  E_z +\nabla \times E_z \right] e^{-i \boldsymbol{k} \cdot r}.\label{appB34}
\end{equation}
\noindent

\end{document}